\documentclass[12pt,preprint]{aastex}

\usepackage{graphicx}
\usepackage{textcomp}

\newcommand{\permil}{\textperthousand}
\newcommand{\delval}[3]{$\delta^{#1}\textrm{#3}/^{#2}\textrm{#3}$}
\newcommand{\ratio}[3]{$^{#1}\textrm{#3}/^{#2}\textrm{#3}$}
\newcommand{\iso}[2]{$^{#1}\textrm{#2}$}
\newcommand{\gcm}{$\textrm{g}/\textrm{cm}^{3}$}
\newcommand{\Msun}{M$_{\odot}$}
\newcommand{\pistar}{$\pi^*$}
\newcommand{\sigstar}{$\sigma^*$}

\newcommand{\angstrom}{\AA{}}

\slugcomment{ApJ}
\shortauthors{Evan Groopman, Larry R. Nittler, Thomas Bernatowicz, Ernst Zinner}

\begin{document}
\title{NanoSIMS, TEM, and XANES studies of a Unique Presolar Supernova Graphite Grain}
\author{Evan Groopman}
\affil{Laboratory for Space Sciences, Physics Department, Washington University}
\affil{One Brookings Drive, Campus Box 1105, Saint Louis, MO 63130}
\email{eegroopm@physics.wustl.edu}
\and
\author{Larry R. Nittler} 
\affil{Department of Terrestrial Magnetism, Carnegie Institution of Washington}
\affil{5241 Broad Branch Road, NW, Washington, DC 20015-1305}
\and
\author{Thomas Bernatowicz, Ernst Zinner}
\affil{Laboratory for Space Sciences, Physics Department, Washington University}
\affil{One Brookings Drive, Campus Box 1105, Saint Louis, MO 63130}

\begin{abstract}
We report on isotopic and microstructural investigations of a unique presolar supernova (SN) graphite grain, referred to as G6, isolated from the Orgueil CI chondrite. G6 contains complex heterogeneities in its isotopic composition and in its microstructure. Nano-scale secondary ion mass spectrometer isotope images of ultramicrotome sections reveal heterogeneities in its C, N, and O isotopic compositions, including anomalous shell-like structures. Transmission electron microscope studies reveal a nanocrystalline core surrounded by a turbostratic graphite mantle, the first reported nanocrystalline core from a low-density SN graphite grain. Electron diffraction analysis shows that the nanocrystalline core consists of randomly oriented 2--4 nm graphene particles, similar to those in cores of high-density (HD) presolar graphite grains from asymptotic giant branch stars. G6's core also exhibits evidence for planar stacking of these graphene nano-sheets with a domain size up to 4.5 nm, which was unobserved in the nanocrystalline cores of HD graphite grains. We also report on X-ray absorption near-edge structure measurements of G6. The complex isotopic- and micro-structure of G6 provides evidence for mixing and/or granular transport in SN ejecta.
\end{abstract}


\section{Introduction}
\label{sec:intro}
\par Presolar grains condensed in the ejecta of late-stage stellar objects such as asymptotic giant branch (AGB) stars and Type-II supernovae (SNe) prior to the formation of the Solar System. These grains survived travel through the interstellar medium, incorporation into the protosolar nebula, the formation of the Solar System, and entrainment in asteroids and comets. Primitive meteorites preserved this stardust, from where the grains may be isolated and studied in the laboratory. Each presolar grain is a stellar fossil of its progenitor star; microstructural and isotopic study allows for inferences regarding the formation conditions and nucleosynthesis processes in the parent star. Many different species of presolar minerals have been discovered \citep{zinner2014} including: nanodiamond \citep{lewis1987}, SiC \citep{bernatowicz1987}, graphite \citep{amari1990}, silicates \citep{messenger2003}, oxides \citep{hutcheon1994,nittler1994}, and carbide subgrains \citep{bernatowicz1991}, which form in various types and/or outflow regions of stars.

\par Presolar graphite grains are generally quasi-spherical \citep{zinner1990,amari1994}; their internal crystal structures range from highly-ordered concentric shells of graphite \citep{bernatowicz1996,croat2005} to more disrupted turbostratic graphite \citep{croat2003,croat-MAPS2008}. Presolar graphite spherules isolated from the Murchison (CM2) and Orgueil (CI1) carbonaceous chondrite meteorites  have been divided into two major populations on the basis of density: high density (HD) ($\rho >$ 2.05\gcm; $\rho$ = 2.09--2.23 \gcm\ for pure graphite) and low density (LD) ($\rho$ = 1.60--2.05 \gcm) \citep{amari1994,jadhav2006}. Isotopic studies indicate that HD and LD grains generally formed in the outflows of AGB stars and SNe, respectively \citep{hoppe1995,zinner1995,bernatowicz1996,jadhav2006}. A relatively small number of HD graphite grains have been identified as having a SN or a Born-Again AGB star origin \citep{jadhav2008,jadhav-GCA-2013,jadhav:apjl2013}, though these grains will not be discussed further here. HD graphite spherules typically have an onion-like internal structure consisting of concentric sheets of well-ordered graphite \citep{bernatowicz1996}. In contrast, LD graphite grains are typically comprised of turbostratic graphite, with shorter, curved, discontinuous sheets, where the average distance between stacked graphene sheets is up to 5\% larger than in ideal graphite \citep{croat-MAPS2008}. Increased disorder in presolar graphite grains has been found to correlate with increased O content \citep{croat-MAPS2008}.  Approximately one third of HD graphite grains have been found to contain nanocrystalline cores consisting of randomly oriented graphene sheets 3--4 nm in size that are surrounded by mantles of graphite \citep{croat2005}. It has been suggested that these nanocrystalline cores formed when C was highly supersaturated in the stellar outflows where they formed, and the transition to longer-range order (e.g., mantling graphite) occurred when the number density of condensable C atoms diminished \citep{bernatowicz1996}. Since nanocrystalline cores are always found within mantles of well-graphitized carbon, and never vice versa or in an alternating pattern, these grains likely reflect monotonically changing physical and/or chemical environments in AGB outflows. Prior to the present work, there have been no reported findings of LD SN graphite grains containing nanocrystalline cores. 
\par OR1d6m, described in \citet{groopman2012}, is the sixth sample mount of LD presolar graphite grains from the OR1d size/density fraction of the Orgueil CI chondrite ($\rho$ = 1.75--1.92\gcm) \citep{jadhav2006}. OR1d6m contains grains with likely SN origins based upon Nano-scale Secondary Ion Mass Spectrometry (NanoSIMS) isotopic measurements. One grain, OR1d6m-6 ($\sim$6.5 \micron\ in diameter, hereafter G6), was selected for further microanalytical and isotopic study in the NanoSIMS, Transmission Electron Microscope (TEM), and Scanning Transmission X-ray Microscope (STXM).

\section{Methods}
\label{sec:methods}
\par After bulk-grain NanoSIMS analysis \citep{groopman2012}, G6 was electrostatically ``picked'' off the mount with an electro-chemically-sharpened W needle and deposited in a drop of LR White hard resin at the bottom of a polyethylene SPI BEEM 1001 capsule. The capsule tapers pyramidally into a 1 mm $\times$ 1 mm square face. Before picking, the face was visually inspected under a microscope to ensure a minimal degree of roughness. Carbon fibers (50 \micron\ $\times$ 10 \micron) were deposited in the drop of resin and arranged around the grain to aid in locating G6 in the TEM and STXM. The capsule's tapered space was then completely filled with resin and cured for 24 hours in a vacuum oven at 70$^\circ$C. Following the curing process, a steel screw was glued onto the base of the resin block with epoxy and the polyethylene capsule was removed. A glass knife was used to trim the square face of the resin block to a 250 \micron\ $\times$ 250 \micron\ square centered on the grain and fibers. The tip of the resin block was then ultramicrotomed into 70 nm-thick sections. The majority of microtomed sections were deposited on holey-C-coated Cu TEM grids for TEM and STXM analyses, while the remainder sections were deposited on a Si wafer for NanoSIMS isotope imaging \citep{groopman_metsoc2013}. The latter were chosen to be from the center region of G6.

\subsection{Secondary Ion Mass Spectrometry}
\label{sec:methodSIMS}
\par Isotopic images of the microtome sections on the Si wafer were obtained with the Cameca NanoSIMS 50L at the Department of Terrestrial Magnetism, Carnegie Institution of Washington. Under primary Cs$^+$ bombardment, secondary ions of \iso{16}{O}$^-$, \iso{18}{O}$^-$, \iso{12}{C}\iso{12}{C}$^-$, \iso{12}{C}\iso{13}{C}$^-$, \iso{12}{C}\iso{14}{N}$^-$, \iso{12}{C}\iso{15}{N}$^-$, and \iso{28}{Si}$^-$ were collected in multicollection mode. C$_2$ and CN signals were used to determine the \iso{12}{C}/\iso{13}{C} and \iso{14}{N}/\iso{15}{N} ratios, respectively. Single ions and molecular ions have different energy distributions, so measuring C$_2$ allows for better alignment with the CN secondary ions \citep{degregorio2013}, which are commonly used for N-isotopic measurements by SIMS \citep{zinner1987}. Additionally, carbonaceous materials often yield ion signals where C$_2^-$/C$^-$ $>$ 1, allowing for higher statistical precision. Isotope images were obtained over 5--15 \micron\ raster areas, depending on the cross section size, of 256 pixels $\times$ 256 pixels with a dwell time of 10 ms/px. Typically 10--20 image frames were gathered per measurement or until the microtome section was sputtered away. Images were analyzed with the custom L'image software (L. R. Nittler, Carnegie Institution). The carbon fibers and resin were measured as standards.

\subsection{TEM}
\label{sec:methodTEM}
\par Microtome sections on TEM grids were analyzed with a JEOL  high-resolution JEM-2100F TEM, and a JEOL 2000FX TEM equipped with a NORAN ultra-thin window energy dispersive X-ray spectrometer (EDXS). Quantitative EDXS analysis was performed on all internal subgrains using the Cliff-Lorimer method \citep{cliff1975}. Electron diffraction patterns were obtained on film in the JEM-2100F and were subsequently digitized. Radial profiles across the digitized diffracted electron intensity images were obtained with a modified version of the open source Diffraction Ring Profiler \citep{zhang2011}. The code was modified to use a polynomial + Lorentzian profile of the form:

\begin{mathletters}
\begin{equation}
c_0+c_1x + \frac{A}{2\pi}\times \frac{w}{(x-x_0)^2+(w/2)^2}
\label{polyLorentz}
\eqnum{1}
\end{equation}
\end{mathletters}

\noindent to model the background diffracted intensity from the central spot, where $c_{0,1}$ are polynomial coefficients, $A$ is the area under the Lorentzian profile, $w$ is the full width at half maximum (FWHM), and $x_0$ is the median/center position. The user selects points on the background profile, after which the background is fitted and subtracted. The polynomial-Lorentzian model provides a near-perfect match to the background diffraction intensity, substantially better than the Diffraction Ring Profiler's built-in pseudo-Voigt and power-law profiles. Individual diffraction peaks often are not symmetric, having tails toward higher reciprocal spacing. This characteristic is due both to instrumental broadening and to the fact that the patterns are obtained from finite, non-ideal crystals. Individual diffraction peaks were therefore fitted to an asymmetric Lorentzian profile (Eq.~\ref{Lasym}), utilizing nonlinear least-square minimization available through the Python package lmfit \citep{lmfit}, where the FWHM, $w$, was allowed to vary sigmoidally according to an asymmetry parameter, $a$ (Eq. \ref{wx}):

\begin{mathletters}
\begin{eqnarray}
L_{sym.} &=& \frac{A}{2\pi}\times \frac{w}{(x-x_0 )^2+(w/2)^2} \label{Lsym}\\
w(x) &=&  \frac{2w_0}{1+e^{-a(x-x_0 )}} \label{wx}\\
L_{asym.} &=& \frac{A}{\pi}\times \frac{w_0/(1+e^{-a(x-x_0 )})}{(x-x_0 )^2+(w_0/(1+e^{-a(x-x_0 )}))^2}
\label{Lasym}
\end{eqnarray}
\end{mathletters}

\noindent adapted from \citet{stancik2008}. Note that when $a\neq$ 0, the peak's median reciprocal space location is no longer identical with the location of the peak's maximum intensity. Additionally, $w_0$ is the full width at half the intensity at the median reciprocal distance, which is larger than the FWHM. To obtain the FWHM, one must substitute the location of the peak maximum into Eq. \ref{wx} or calculate it numerically from the fit. When $a$ = 0, $w$ becomes a constant and the Lorentzian returns to its symmetric form (Eq. \ref{Lsym}). This form of asymmetric Lorentzian is advantageous because it models peak tails very well and $w$ is constrained between the values 0 and 2$w_0$. Where two or more diffraction intensity peaks overlapped, all were fitted simultaneously, however only the largest peak was modeled as an asymmetric Lorentzian; the other peak(s) were fitted to symmetric Lorentzian profiles to reduce the number of fitting parameters.

\par
\begin{figure}[ht]
\figurenum{1}
\plotone{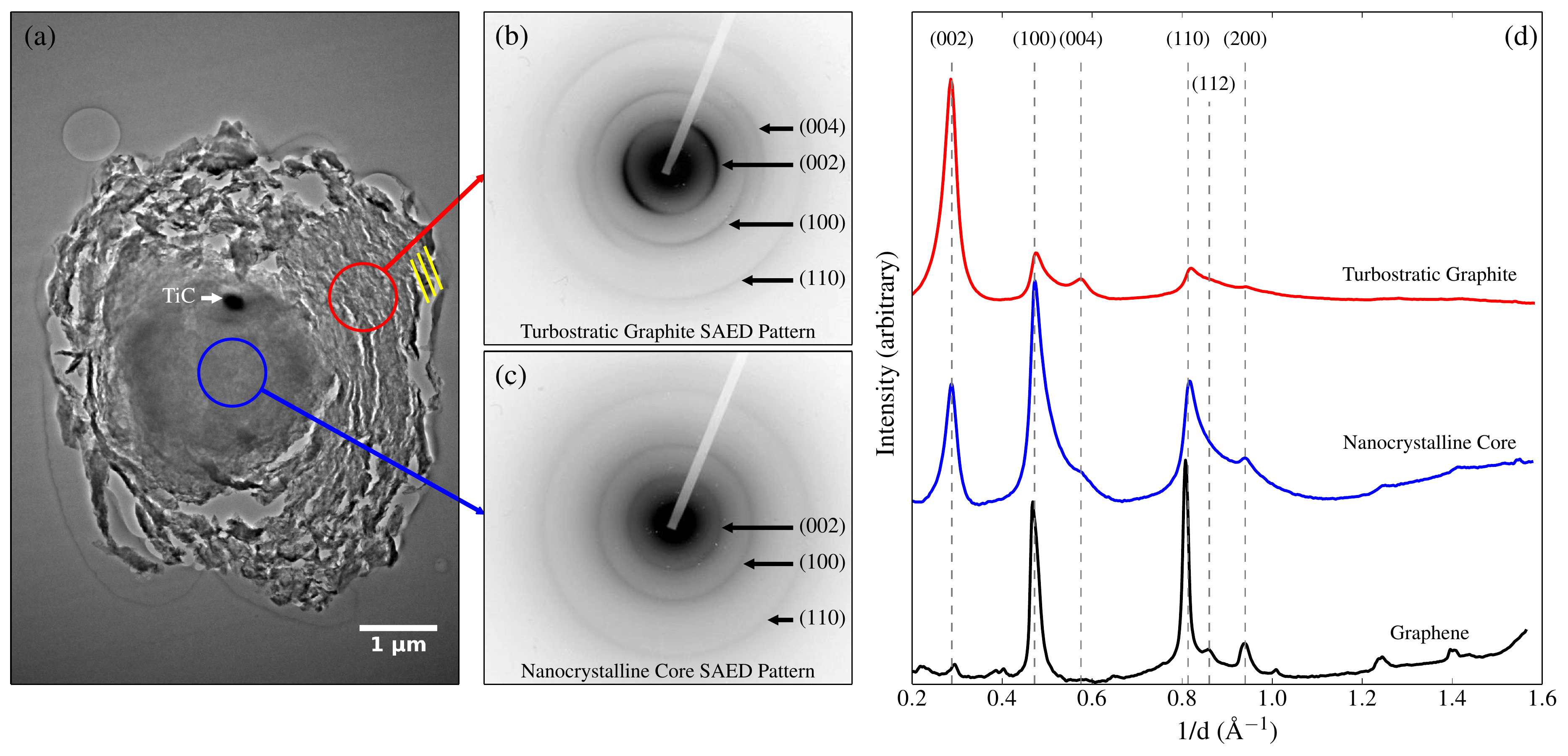}
\epsscale{.9}
\caption{\singlespace(a) TEM bright field (BF) micrograph of an ultramicrotome section of presolar graphite grain G6 containing a nanocrystalline core surrounded by turbostratic graphite. It is easy to distinguish between the two regions by visual inspection. A 265 nm TiC subgrain (darkly contrasting) is visible at the edge of the core region. Circles indicate regions where SAED patterns (b--c) were obtained on film with 60s exposures. Parallel lines indicate edge-on orientation of local graphitic planar stacking. (b) SAED pattern of the turbostratic graphite in reverse contrast for easier viewing. Diffracted intensity for (002) and (004) reflections is highest when planar stacking is parallel to the electron beam. Diffracted intensity for in-plane (100) and (110) reflections is largest orthogonal to the largest intensity for planar stacking. (c) SAED pattern of nanocrystalline core. All diffraction rings show no orientational bias, indicating that the core consists of randomly-oriented crystals. SAED pattern includes previously unobserved, but weak (002) reflection, indicating that crystals inside of the core exhibit some degree of planar stacking. (d) Diffraction ring profiles of (b--c), and graphene for calibration (SAED not shown). The graphene was ultrasonically-dispersed from Ceylon graphite and therefore consists of single- and multi-layered regions, hence the presence of a small (002) reflection. Individual profiles are normalized for visual effect; SAED patterns show the absolute intensities. The turbostratic graphite profile is dominated by the strong (002) reflection. The most intense reflection for the nanocrystalline core is the (100) in-plane reflection, comparable in intensity to the (100) reflection from turbostratic graphite (as seen on film), though a weaker (002) reflection is present. Based upon the (100) and (110) peak widths we estimate the average lateral particle size to be 2--4 nm, with a size of $\sim$4.5 nm in the (002) stacking direction.}
\label{fig:Ediff}
\end{figure}

 Ceylon graphite was used as a calibration standard to determine instrumental broadening of electron diffraction peaks. Flakes of Ceylon graphite were ultrasonically dispersed in distilled water and the suspension was deposited on TEM grids. The invariant in-plane (100) and (110) diffracted intensity peaks were fitted to symmetric Lorentzian profiles to determine the FWHM and d-spacings. Due to the ultrasonic dispersion, most of the deposited Ceylon graphite was actually single- or few-layered graphene, i.e. there is little to no planar stacking. Therefore, in electron diffraction patterns the (002) reflection is nearly nonexistent; in Figure~\ref{fig:Ediff} this ring profile is labeled as ``graphene''.

\subsection{STXM}
\label{sec:methodSTXM}
\par After TEM analyses, sections of G6 were analyzed with the synchrotron-based STXM at Beamline 5.3.2.2 of the Advanced Light Source, Lawrence-Berkeley National Laboratory \citep{kilcoyne2003}. Monochromatic X-rays are generated by electrons transmitted through a dedicated bending-magnet, followed by a spherical-grating monochromator that allows for selection of photon energy up to 800 eV in steps as small as 0.1 eV. The beam is focused by Fresnel zone plate optics into a 25--40 nm spot size at the sample. A three-dimensional data ``stack'' in x $\times$ y $\times$ eV was generated by using an interferometer-controlled piezoelectric stage to raster the sample across the beam, and by stepping the energy up after each frame. The photon energy was increased in steps of 0.1 eV between 283 eV and 296 eV, across the C K-edge, while 0.5 eV steps were used in the pre- and post-edge regions due to time constraints. The absorbance, or optical density (OD), of the sample was calculated as $-\log(I/I_0)$, where $I$ was the measured intensity through the sample and $I_0$ was the baseline intensity measured through a hole in the sample. Each pixel within the stack contains an X-ray absorption near-edge structure (XANES) spectrum.
\par Within the C K-edge region there are three diagnostic absorption features for graphite and aromatic carbon: a peak at 285.2 eV, corresponding to the 1s$\rightarrow$2p \pistar\ transition; a sharp peak at 291.5 eV, corresponding to the \sigstar\ exciton; and a broader \sigstar\ peak at 292.5 eV \citep{bruhwiler1995,ahuja1996,brandes2008}. The \pistar\ peaks were fitted using the same asymmetric Lorentzian profile described above for TEM diffraction patterns, which performs well in modeling the tail present above the peak absorption energy. Prior to fitting the \pistar\ peaks, a linear fit to the pre-edge region between 270 -- 282 eV was subtracted from the spectra. The \pistar\ peaks were fitted without modeling the C K-edge as an arctangent or step function. The aim was to compare the \pistar\ peaks by utilizing as few parameters as possible. Additionally, ionization potentials for the relevant C bonds lie far enough above the aromatic \pistar\ peak \citep{cody1995}, that they need not be subtracted from the spectrum to quantify the \pistar\ resonance. Peak positions were calibrated using CO$_2$ gas prior to any measurements on G6.

\par Spectra from organic carbonaceous materials often contain other minor peaks in the region between the aromatic \pistar\ and \sigstar\ excitons, including those associated with 1s-\pistar\ transitions in ketones (286.5 eV), aliphatics (287.3 -- 288.1 eV), and carboxyl (288.4 -- 288.7 eV) \citep{codyMAPS2008,degregorio2013}. To fully deconvolve the spectra and quantify any minor peaks, a continuum step ionization potential for aromatic carbon was subtracted from the spectra. The continuum step was modeled as an error function (erf) of the form:

\begin{mathletters}
\begin{equation}
I = H\Big[\frac{1}{2} + \frac{1}{2}\textrm{erf}\Big(\frac{E-P}{2\textrm{ln}(2)\times \Gamma}\Big)\Big]
\label{eqn:step}
\eqnum{3}
\end{equation}
\end{mathletters}

\begin{figure}[ht]
\figurenum{2}
\includegraphics[width=0.5\textwidth]{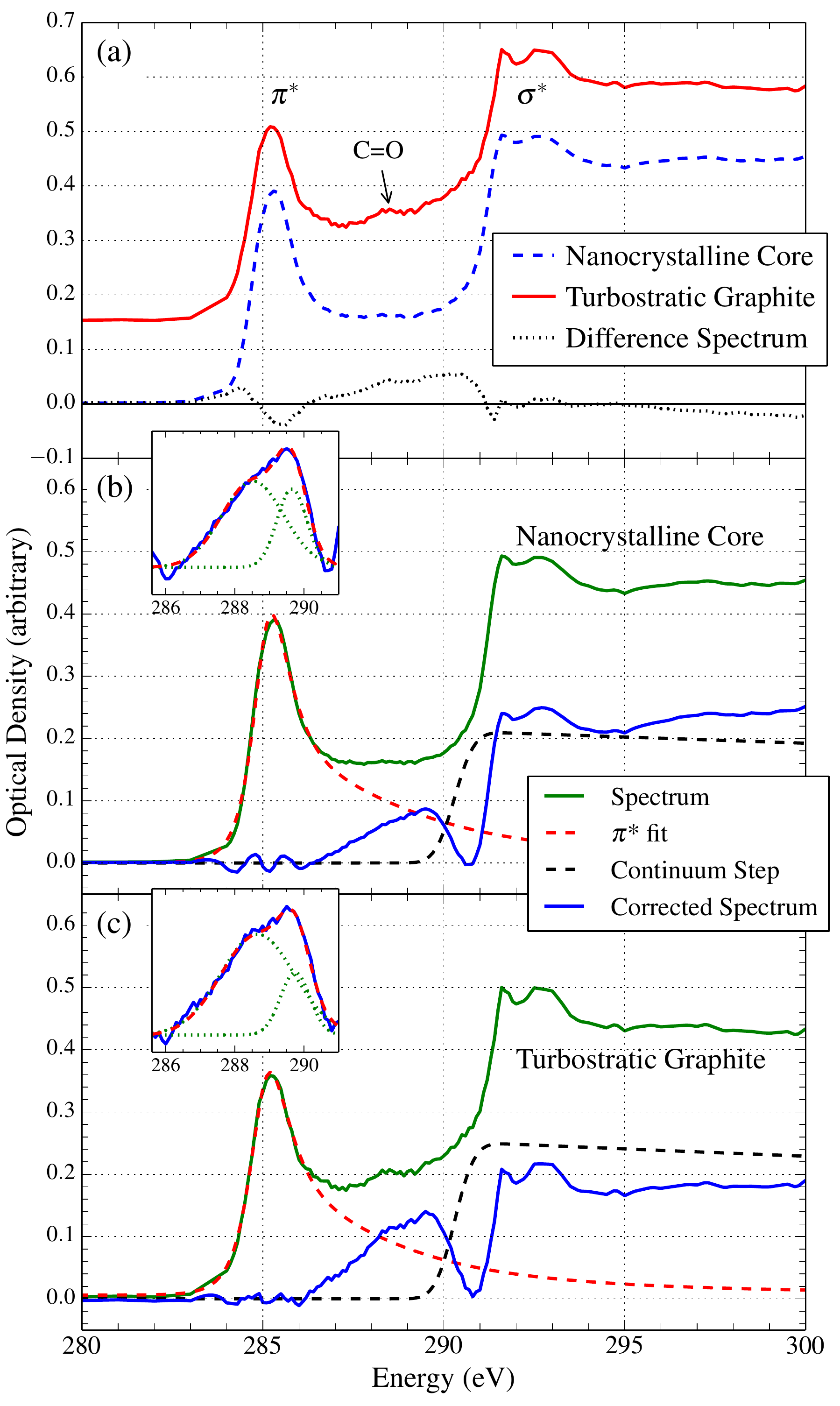}
\caption{\singlespace(a) C K-edge XANES spectra from the nanocrystalline core and mantling turbostratic graphite of presolar grain G6. Both the core and mantle exhibit graphite-like spectra, with aromatic 1s-\pistar$_1$ transitions at 285.3 eV and 285.4 eV, respectively, and \sigstar\ transitions both at 291.5 eV and 292.5 eV. The spectra are normalized to the area under the \sigstar\ peaks. The primary difference between the spectra is the presence of a minor peak matching carboxyl in the turbostratic graphite spectrum. (b-c) Deconvolution of spectra from nanocrystalline core and turbostratic graphite, respectively. Insets show deconvolution of minor peaks into two Gaussian profiles. The smaller, higher-energy peak corresponds to the 1s-\pistar$_2$ transition from conjugated $\pi$ bonds. The peak at $\sim$288.6 eV is twice as intense in the turbostratic graphite spectrum than in the spectrum from the nanocrystalline core, while the \pistar$_2$ peaks are identical. See the electronic edition for a color version of this figure.}
\label{fig:XANES}
\end{figure}

\noindent\citep{stohr1992} where $E$ is the energy, H is the height of the step, $P$ is the ionization potential (290.3 eV for aromatic carbon \citep{cody1995}), and $\Gamma$ is the width of the step in eV. At energies higher than the ionization potential, the continuum step decays exponentially \citep{stohr1992}, however this does not affect the deconvolution of minor peaks below the ionization potential. $\Gamma$ is intrinsically linked to the instrumental resolution \citep{stohr1992} and has been chosen to be 1 eV in earlier work (e.g. \citet{cody1995}). Deconvolutions of the nanocrystalline (core) and turbostratic graphite (mantle) spectra (see section \ref{sec:nanocore}) are illustrated in Figures \ref{fig:XANES}b-c, where the spectra have been normalized to the area under the \sigstar\ peaks. The asymmetric Lorentzian profiles described above (dashed red in the online version) and continuum steps located at 290.3 eV with widths of 1 eV (dashed black) are subtracted from the absorbance spectra (solid green online). Continuum step heights of 0.21 and 0.25 for the core and mantle, respectively, were chosen so that the residual spectrum (solid blue online) was minimized and non-negative. Graphite is composed of conjugated C, so the $\pi$ molecular orbitals are delocalized above and below the graphene sheets, which results in a splitting of the main \pistar\ peak into a large component at 285.2 eV (\pistar$_1$)  and a smaller component (\pistar$_2$) $\sim$4 eV higher in energy \citep{stohr1992,cody1995}.  The insets of Figures \ref{fig:XANES}b-c show a magnified view of the minor peak region where two Gaussian profiles (dotted green online) were fitted simultaneously (dashed red online) to the residual spectrum. In both insets the smaller of the two peaks at higher energy is the \pistar$_2$ peak. Given the contribution from the tail of the asymmetric Lorentzian profile fit to \pistar$_1$, the areas under the minor peaks can be assumed to be lower limits. The drawback of this method is that the \pistar\ peak and the minor peaks are not fitted simultaneously during the deconvolution, however we do not overestimate the ionization potential's contribution, given the asymmetric model of the \pistar\ peak. 

\section{Results and Discussion}
\label{sec:Results}
\subsection{Isotopic Composition}
\label{sec:IsoComp}
\par
\begin{figure}[ht]
\figurenum{3}
\plotone{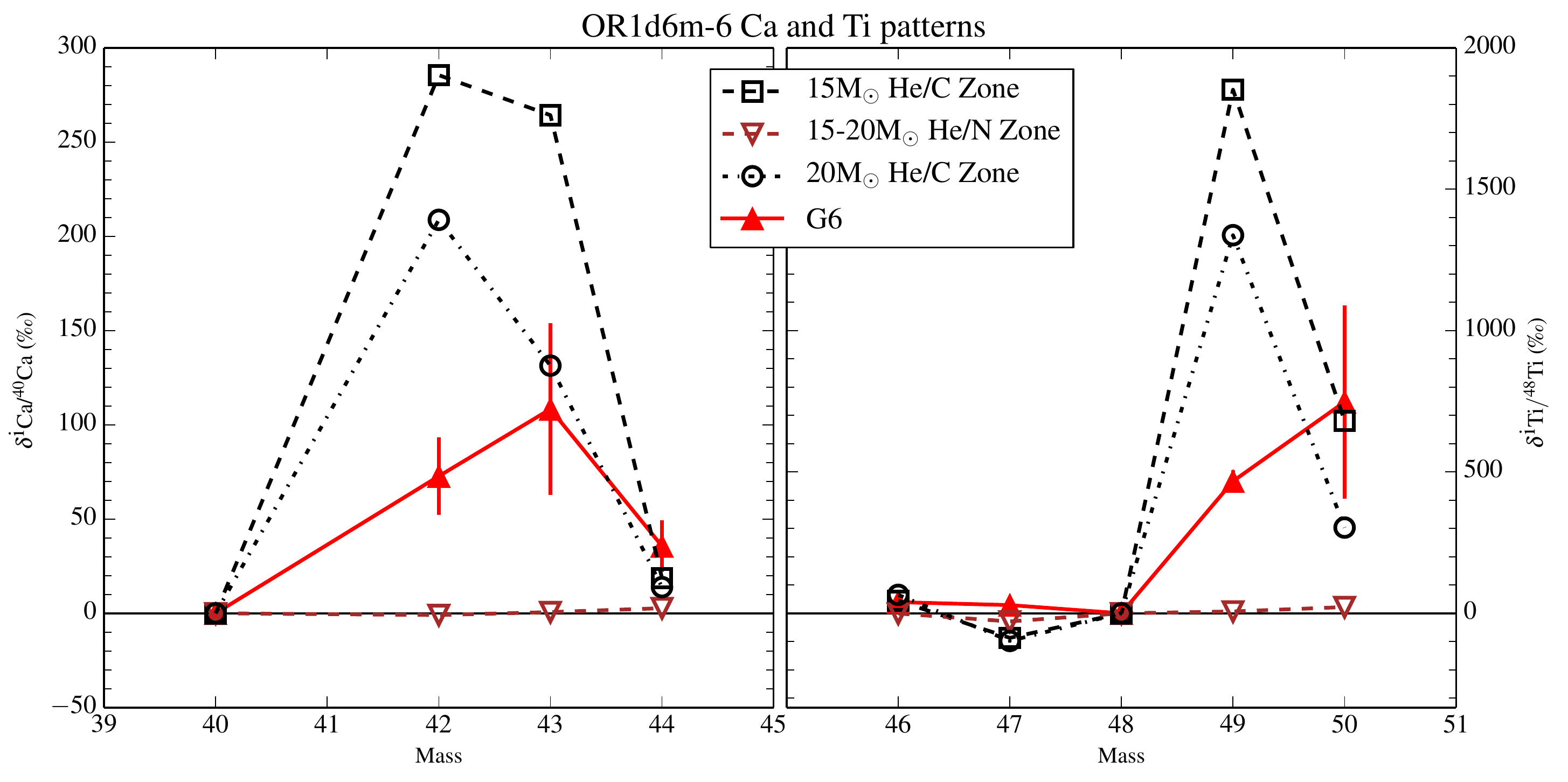}
\caption{\singlespace Ca and Ti isotope patterns (solid triangles, 1-$\sigma$ errors) for presolar graphite grain G6. Dashed and dotted lines are average compositions from He/C and He/N zones in 15 and 20 \Msun\ Type-II SN models \citep{rauscher2002}. G6 exhibits typical neutron capture patterns in its Ca and Ti isotopic compositions. The relatively large error bars on the Ca $\delta$-values are due to its low abundance. The large error bar on $\delta$\iso{50}{Ti} is due to isobaric interference from \iso{50}{Cr}; the Cr signal is dominated by contamination as a result of the acid separation process and resulted in a large correction on \iso{50}{Ti}. The isotope patterns for both elements can be matched by the composition of a mixture of material from the He/C and He/N zones.}
\label{fig:CaTi}
\end{figure}

 Bulk isotopic measurements of G6 were made as a part of the analysis of the OR1d6m suite of LD graphite grains from Orgueil \citep{groopman2012}. A SN origin for G6 is indicated by large excesses in \iso{12}{C}, \iso{15}{N}, and \iso{18}{O} relative to solar system isotopic compositions; a small excess in \iso{28}{Si}; and a large \iso{26}{Mg} excess indicating a very high initial abundance of  short-lived (t$_{1/2}$ = 7.2$\times10^5$ yrs.) \iso{26}{Al} (Table \ref{table:G6}). Presupernova massive stars consist of concentric shells of material defined by their most abundant elements and the different stages of nucleosynthesis which they experience \citep{woosley1995,meyer1995} (see Figure 9 of \citet{zinner2014}). Of particular note to this study are three zones: the He/C zone, characterized by triple-$\alpha$ burning to \iso{12}{C} and the \iso{14}{N}($\alpha$,$\gamma$)\iso{18}{O} reaction; the He/N zone, characterized by the CNO cycle and a prodigious producer of \iso{26}{Al} via hydrogen burning; and the Si/S zone, where \iso{28}{Si} is abundantly produced via O-burning: \iso{16}{O} + \iso{16}{O} = \iso{28}{Si} + \iso{4}{He}. Presolar grains from SNe, including SiC X grains, silicon nitride (Si$_3$N$_4$) grains, and LD graphite grains, often exhibit nucleosynthetic signatures from multiple zones, which implies that mixing in the ejecta must occur \citep{travaglio1999}. In isotopic surface images of G6, spatially-correlated extreme excesses in \iso{18}{O} and \iso{15}{N} were found, with \delval{18}{16}{O}\footnote{\delval{i}{j}{X} = $1000 \times$ \big((\ratio{i}{j}{X})$_{\textrm{sample}}$/(\ratio{i}{j}{X})$_{\textrm{standard}} - 1\big)$} up to 98,000\permil\ (\ratio{16}{18}{O} = 5; terrestrial = 499) and \delval{15}{14}{N} up to 6,400\permil\ (\ratio{14}{15}{N} = 37; terrestrial = 272) (see Figure 1 of \citet{groopman2012}). The extreme O and N isotopic ratios of these surface hotspots are much closer to the ratios of the He/C zone in models of 15--20 \Msun\ SNe \citep{rauscher2002} than the ratios in the rest of the grain, which likely resulted from a mixture also including material from the overlying He/N zone. Spatially-correlated hotspots of excesses in \iso{15}{N} and \iso{18}{O} have in some cases been found to be carried by internal TiC subgrains \citep{groopman2012}, which are indicative of material from the inner portion of the He/C zone \citep{rauscher2002,bojazi2014}. In addition to excesses in \iso{15}{N} and \iso{18}{O}, excesses in \iso{12}{C}, \iso{42,43,44}{Ca}, and \iso{49,50}{Ti} are also signatures of the He/C zone in models of 15--20 \Msun\ SNe, though once more G6's measured isotopic compositions are less anomalous than those predicted for pure He/C zone material (Table \ref{table:G6} and Figure \ref{fig:CaTi}). The relatively large errors in Ca isotope enrichments are due to the low abundance of Ca in G6. The large error in \iso{50}{Ti} is due to isobaric interference from \iso{50}{Cr}. The Cr$^+$ signal for G6 is large and overwhelmingly due to dichromate contamination from the acid dissolution of the meteorite \citep{amari1994,jadhav2006}. The inferred \iso{26}{Al}/\iso{27}{Al} ratio for G6 is one of the highest ever observed in a presolar graphite grain and is indicative of material from the He/N zone, whose \iso{26}{Al}/\iso{27}{Al} ratio is 10$\times$ higher than that of the He/C zone. G6's \iso{26}{Al}/\iso{27}{Al} ratio, however, is even larger than the average predicted ratio in the He/N zone for 15--20 \Msun\ SNe. Excess \iso{28}{Si} points to a contribution from the deep Si/S zone, however this excess is marginal in G6. Combined, G6's isotopic composition matches that of a mix of material predominantly from the He/C and He/N zones, which also have C/O $>$ 1, a condition favorable for condensation of carbonaceous grains, with a small contribution from the Si/S zone.

\par
\begin{figure}[ht]
\figurenum{4}
\plotone{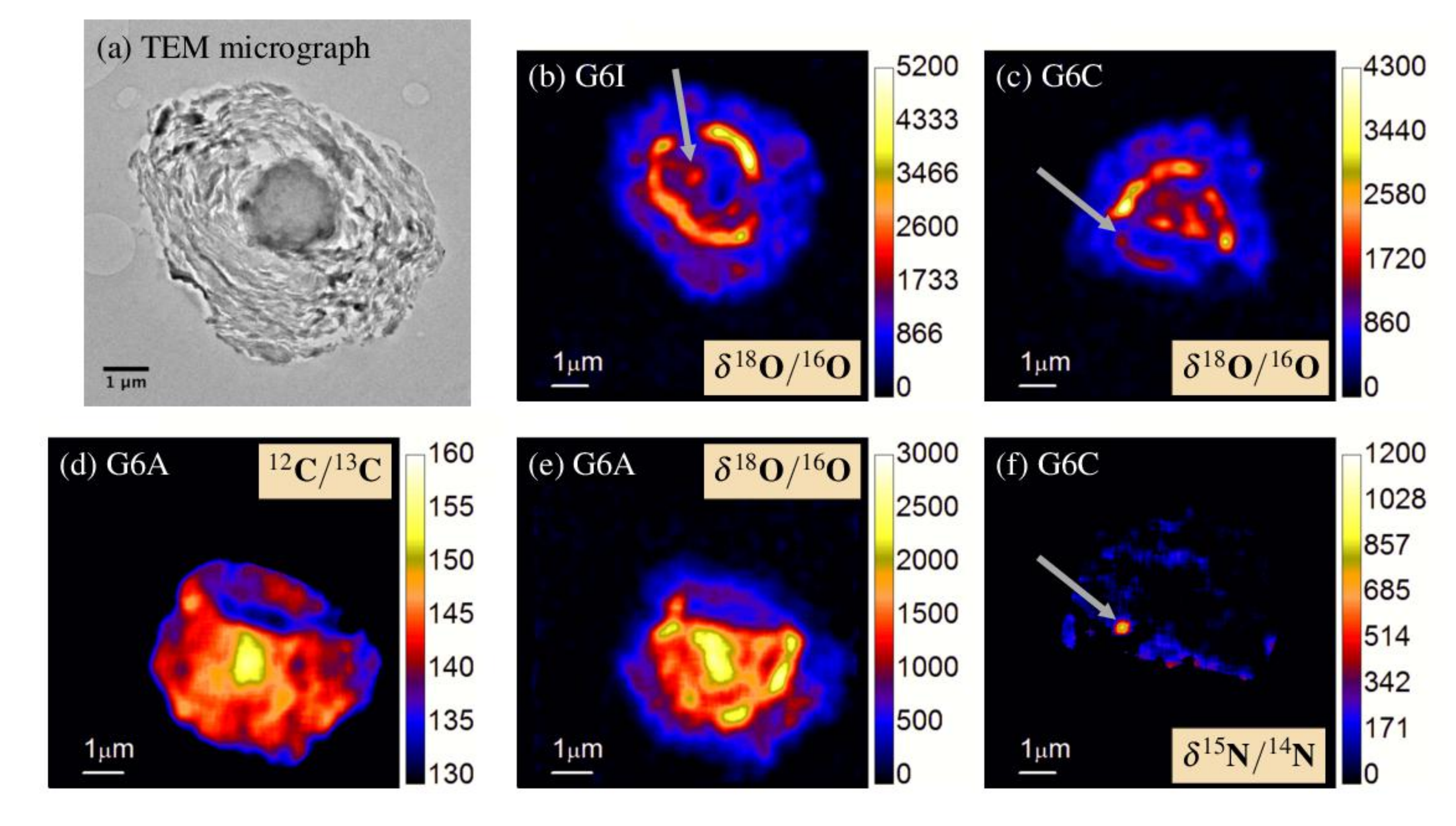}
\caption{\singlespace(a) TEM micrograph of an ultramicrotomed section of G6 containing a nanocrystalline core (1.5 \micron) surrounded by turbostratic graphite. (b--f) Isotope ratio images of three different ultramicrotome sections on a Si wafer. Most notable are the complex \delval{18}{16}{O} anomalies in (b,c,e), including ring-like structures $\sim$2.5 \micron\ in diameter. The ring-like structures are likely a primary product of a non-monotonically changing formation environment. (b,c,f) Hotspots with correlated excesses in \iso{18}{O} and \iso{15}{N}, indicative of the presence of TiC subgrains (indicated by arrows, see Figure~\ref{fig:isoprofs}c for G6I \delval{15}{14}{N} data). (d,e) Radial gradients in the C and O isotopic compositions (see Figure~\ref{fig:isoprofs}).}
\label{fig:isoimgs}
\end{figure}

\begin{figure}[ht]
\figurenum{5}
\plotone{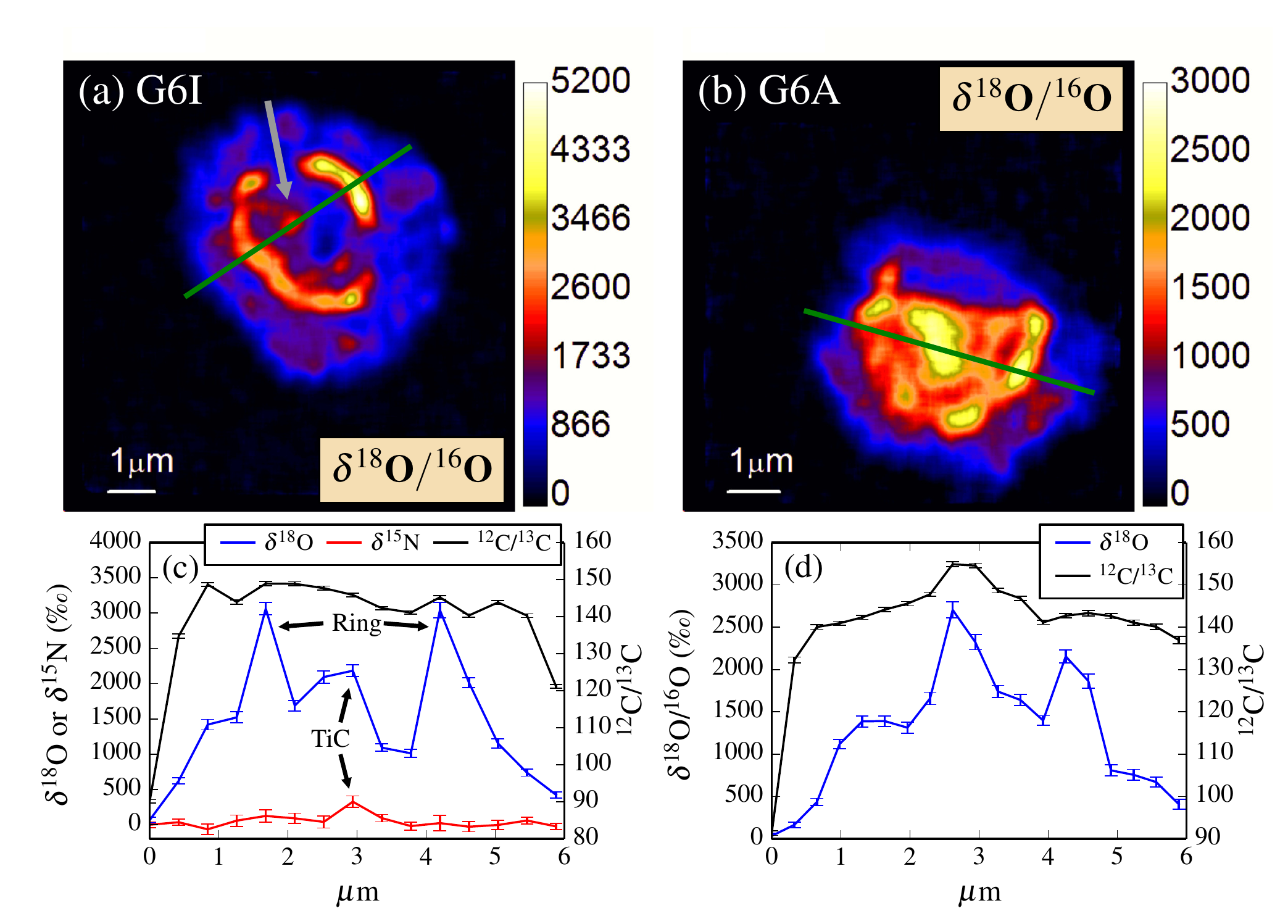}
\caption{\singlespace(a) \& (b) O isotope ratio images of microtome sections G6I and G6A, respectively, of presolar graphite grain G6. (c) \& (d) Isotopic line profiles of sections in (a) \& (b) (green lines). (c) Shows a correlated excesses in \iso{18}{O} and \iso{15}{N} at 2.9 \micron\ corresponding to a TiC subgrain (indicated by arrow in (a)). (d) Radial gradients in C and O isotopic compositions.}
\label{fig:isoprofs}
\end{figure}

 In this work, three different microtome sections (G6A, 4.7 \micron; G6C, 5.7 \micron; G6I, 5.9 \micron) on the Si wafer were isotopically imaged in the NanoSIMS. As seen in the prior bulk grain measurement, G6I is highly enriched in \iso{18}{O} and \iso{12}{C}, however this section is only slightly enriched in \iso{15}{N}. The average isotopic compositions of G6A and G6C are slightly less anomalous than those of G6I. All sections are more anomalous in their average C isotopic compositions than the bulk grain, however the reverse is true for their N and O isotopic compositions. G6A, G6C, and G6I also contain large heterogeneities in their C, N, and O isotopic compositions, including regions whose C and N compositions are much more anomalous than in the bulk-grain measurements. These heterogeneities include both radial gradients in the sections' C and O isotopic compositions (Figures \ref{fig:isoimgs}d-e, \ref{fig:isoprofs}a-b), similar to those observed previously in other SN graphite grains \citep{stadermann2005,groopman2012}, as well as small spatially-correlated hotspots of excesses in \iso{15}{N} and \iso{18}{O} likely corresponding to TiC subgrains (Figures \ref{fig:isoimgs}b,c,f). G6A exhibits a smooth radial gradient in \ratio{12}{13}{C} and a more irregular gradient in \delval{18}{16}{O}, with \ratio{12}{13}{C} ranging from 162 to 147 and \delval{18}{16}{O} from 3000 to 510 (\ratio{16}{18}{O} from 125 to 330) moving from the section's center to its edge (Figure \ref{fig:isoprofs}b). G6C and G6I each contain a $\sim$300 nm hotspot of excesses in \iso{15}{N} and \iso{18}{O} (indicated by arrows in Figures \ref{fig:isoimgs}b,c,f and \ref{fig:isoprofs}a), which correspond to the sizes and locations of TiC subgrains identified in the TEM (section \ref{sec:subgrains}). The G6C hotspot has a composition of \delval{18}{16}{O} = 1840 $\pm$ 82\permil\ (\iso{16}{O}/\iso{18}{O} = 176 $\pm$ 5), and \delval{15}{14}{N} = 1140 $\pm$ 138\permil\ (\ratio{14}{15}{N} = 127 $\pm$ 9); the G6I hotspot \delval{18}{16}{O} = 2590 $\pm$ 93\permil\ (\ratio{16}{18}{O} = 139 $\pm$ 5), and \delval{15}{14}{N} = 430 $\pm$ 86\permil\ (\ratio{14}{15}{N} = 190 $\pm$ 12). All of these values are lower bounds on the isotopic anomalies for the hotspots, as the ion signals for such small regions are diluted by sampling of the surrounding graphite. \citet{groopman2012} showed that the surface of G6 was isotopically heterogeneous, with hotspots in \iso{15}{N} and \iso{18}{O}. These hotspots bias the original bulk grain measurement towards the composition of the presumed TiC subgrains and away from the composition of the less-anomalous graphitic material. Hence, the whole-grain N and O measurements closely match the isotopic composition of the TiC subgrains identified in the microtome sections. The most striking feature of the microtome section isotopic images is the complex O-isotopic heterogeneities in G6C and G6I, notably an \iso{18}{O}-rich ring of material 2.7 \micron\ $\times$ 3.3 \micron\ in size (Figures \ref{fig:isoimgs}b,c, \ref{fig:isoprofs}a). This ring lies outside of G6's nanocrystalline core (1.5 \micron, section \ref{sec:nanocore}), and is more anomalous than material at smaller and larger radii. The ring structure has \delval{18}{16}{O} = 2100 $\pm$ 15\permil\ (\ratio{16}{18}{O} = 161 $\pm$ 1), with little variation in its C and N isotopic compositions relative to the rest of the section. Since G6A is a smaller section than G6C and G6I, and therefore further from the center of the grain, it is likely that the gradient observed in \ratio{16}{18}{O} in this section is simply a cross-section of the O-anomalous shell. The ring features in G6C and G6I cover 2.6 and 2.3 \micron$^2$ out of 25.7 and 27.0 \micron$^2$, respectively. The center of microtome section G6C, 1 \micron\ $\times$ 1.5 \micron\ in size, is also more \iso{18}{O}-enriched than the outer region of the grain (Figure \ref{fig:isoimgs}c) with \delval{18}{16}{O} = 1585 $\pm$ 17\permil\ (\ratio{16}{18}{O} = 193 $\pm$ 1). The dimensions of grain section G6C and of its O-anomalous center are similar to those of the TEM sections which contain a cross section of the nanocrystalline core, though it cannot be unambiguously determined whether the core is present in this particular section. The center of G6I contains a hole near the identified TiC subgrain due to microtome damage, so the O isotopic composition of this region is unknown. It is unknown why the surface hotspots of excesses in \iso{15}{N} and \iso{18}{O} are so much more extreme than those found in G6's interior. In addition, it is interesting that the most \iso{18}{O}-anomalous regions within G6's interior are not carried by TiC subgrains, but are intrinsic features of the graphite structure.

\par Radial gradients in isotopic composition have been attributed to secondary processes, such as isotopic equilibration with the grain's environment during its lifetime (e.g. in the protosolar nebula or in the meteorite parent body) \citep{stadermann2005}, and to primary processes, such as the changing isotopic composition of the stellar ejecta during grain formation \citep{groopman2012}. Changes in the grain's local formation environment could be due either to mixing in the stellar ejecta, and/or to the grain's trajectory through regions with varying isotopic composition \citep{bernatowicz2005}. While isotopic equilibration of minor elements is a significant issue in HD presolar graphite grains \citep{hoppe1995,jadhav2006,jadhav2008,zinner2014},  LD grains typically retain large anomalies in N and O \citep{croat2003,stadermann2005,croat-MAPS2008}. The observations of radial gradients in the C isotopic compositions of LD graphite grains suggest that isotopic equilibration is less prevalent than the minor isotope dilution observed in HD graphite grains \citep{groopman2012}. The turbostratic structure of LD graphite grains would presumably have to be disrupted in order to exchange a significant portion of the C with the surrounding environment. Additionally, the complex structure of the O anomalies in G6 suggests that isotopic equilibration is not the root cause, as isotopic diffusion should produce a smooth radial gradient. If a changing local isotopic environment is the cause for the \iso{18}{O}-enriched ring, this would require that the isotopic composition changed non-monotonically.

\par Recent three-dimensional simulations of Type-II SN explosions, based upon observations of SN 1987A, provide evidence of the large-scale mixing (fractions of AU) that occurs between inner and outer pre-SN zone material in the ejecta immediately following the explosion \citep{hammer2010}. In these models, material from inner zones is ejected asymmetrically through the outer zones, carrying with it a mix of heavy elements in addition to 50--70\% H and He. There are considerable spatial variations in composition within individual ejecta. Ni-Fe-Si and O-Ne-Mg ``bullets'' of material are separately ejected through the C-, He-, and H-rich outer zones, with the heavier knots of Fe-group materials, Ne, and Mg overtaking the O-rich knots, in effect partially inverting the pre-SN concentric shell structure. While such simulations cannot yet provide the resolution required to argue for grain-size-scale mixing, they do provide evidence that stellar ejecta varies spatially in isotopic/chemical composition. The Cassiopeia A SN remnant has also been identified as having resulted from an intrinsically asymmetric explosion, which caused the composition of ejecta to vary spatially \citep{rest2011}. It has been shown for AGB stars that grains are coupled to the radiation field of the star, which accelerates them through the gas via radiation pressure \citep{bernatowicz2005}. In effect, the grains act as miniature sails, being propelled faster than the surrounding gas, causing collisions with the gas in front of them. This could also be achieved from the reverse shock wave from a SN. With further simulations and observations of SNe, there exists the possibility for calculating grain condensation parameters - formation times, radii, velocities relative to gas - as done for AGB stars. More simply, these models provide evidence for the mixing of material between adjacent and non-adjacent SN zones. We do, however, remain far from the goal of connecting the isotopic signatures of pre-SN zones with the composition of presolar grains via high-spatial-resolution mixing models.

\par Isotope ratio images of microtome sections of G6 reveal numerous isotopic heterogeneities. These heterogeneities include spatially-correlated hotspots of excesses in \iso{18}{O} and \iso{15}{N}, which are likely carried by internal TiC subgrains \citep{stadermann2005,groopman2012}. This isotopic signature is indicative of material from the inner He/C zone in Type-II SN \citep{rauscher2002,bojazi2014}, while the bulk grain isotopic composition is more similar to a mixture of material from the He/C and He/N zones \citep{rauscher2002}. The presence of radial gradients in \ratio{12}{13}{C}, and \iso{18}{O}-anomalous shell-like structures provide evidence for mixing in the SN ejecta. G6 likely condensed from a gas that changed isotopically in a non-monotonic fashion.

\subsection{Microstructure}
\label{sec:TEM}
\par The majority of microtomed sections of G6 were deposited on holey-C-coated Cu TEM grids. Some 80\% of the grain's volume is visible in the TEM; the remaining sections were deposited on Si wafers, overlie TEM grid bars, or lie outside the TEM's field of view. G6 is primarily composed of turbostratic graphite, characteristic of LD graphite grains from SNe \citep{croat2003,croat-MAPS2008}.

\subsubsection{Nanocrystalline Core}
\label{sec:nanocore}
\par
\begin{figure}[ht]
\figurenum{6}
\plotone{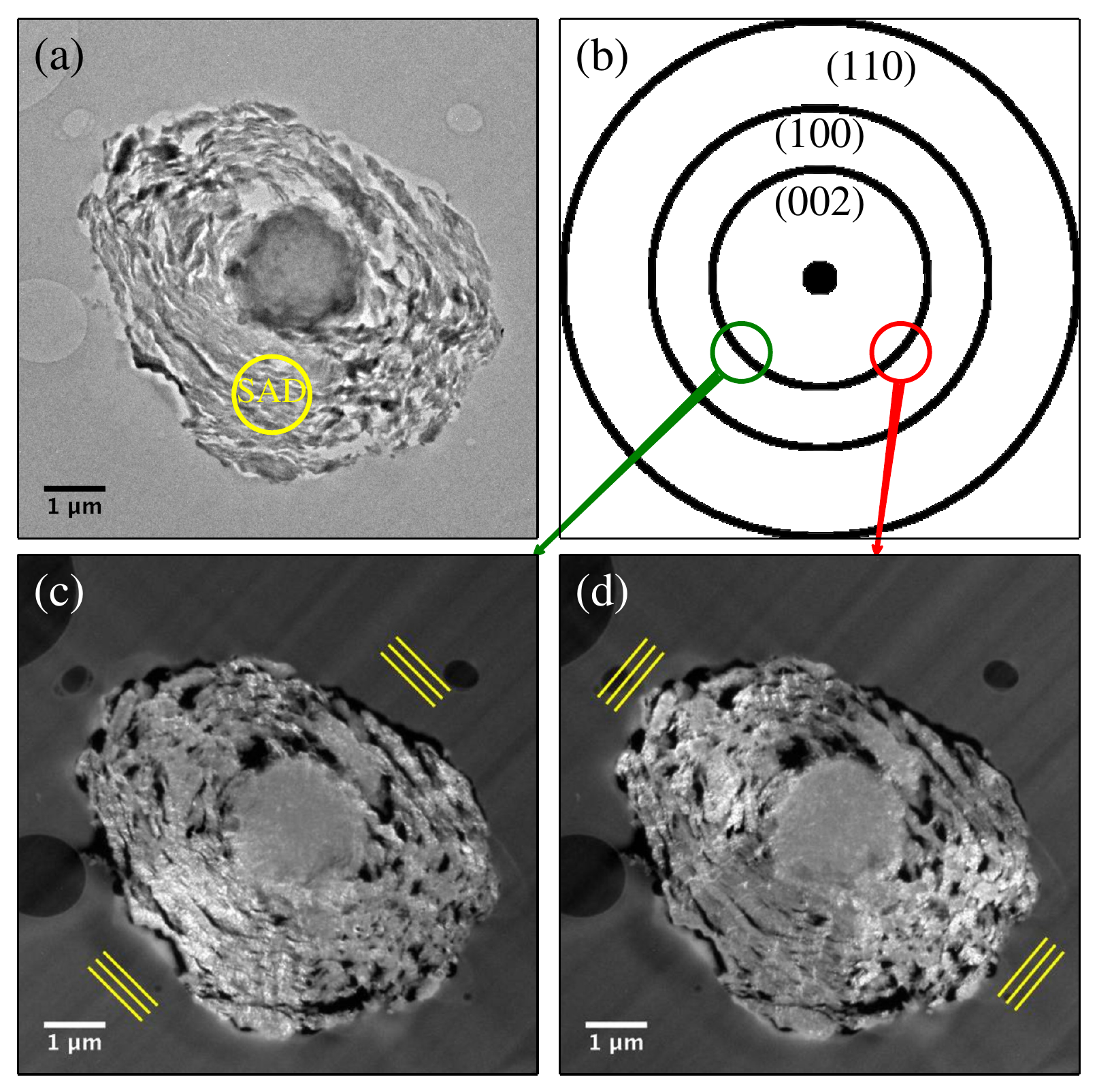}
\caption{\singlespace(a) TEM bright field (BF) image of slice of presolar graphite grain G6. (b) A schematic electron diffraction pattern (intensities and radii not to scale) from a selected area of turbostratic graphite shown in (a). (c) and (d) are dark field (DF) TEM images generated by diffracted electrons from the (002) reflection, selected by an objective aperture at different angles (circles in (b)). The resulting regions of high intensity in the DF images are those where edge-on planar stacking of graphite is parallel to the selected angle of the electron beam (shown with parallel lines). Regardless of the angle selected for generating DF images, the core region remains uniformly intense, indicating an absence of long-range stacking order. Small bright spots are visible within the core, which may be due to larger nanocrystals (10s of nm) within the core; a distribution of these “larger” nanocrystals may be responsible for the (002) reflections observed in core diffraction patterns (Figures \ref{fig:Ediff}c--d). The intensity of the core region is on average greater than that of turbostratic graphite orthogonal to the electron beam, implying that there exists some component of (002) stacking in small particles throughout the core.}
\label{fig:DF}
\end{figure}

 Three sections were found to contain a 1.5 to 2 \micron\ diameter core of nanocrystalline carbon surrounded by a mantle of turbostratic graphite. All other sections are comprised solely of turbostratic graphite. Nanocrystalline cores have only been reported previously from HD onion-type graphite grains from AGB stars \citep{bernatowicz1996,croat2005}.The nanocrystalline and turbostratic regions of G6 are visually distinct in the TEM (e.g., Figs. \ref{fig:Ediff}a; \ref{fig:DF}a,c,d; \ref{fig:graphitesub}a--d). Selected area electron diffraction (SAED) patterns from the turbostratic graphite mantle (Figure \ref{fig:Ediff}b) closely match pure graphite. The dominant feature in the patterns is the (002) reflection (d$_{(002)}^{\textrm{ideal}}$ = 3.354 \angstrom), which corresponds to planar stacking of graphene sheets. The measured (002) d-spacing for turbostratic graphite is 4.9\% larger than that of an ideal crystal (d$_{(002)}^{\textrm{turbo}}$ = 3.52 \angstrom), similar to previous observations of presolar graphite \citep{bernatowicz1996,croat2005}. This is corroborated by the measured d-spacing of the (004) reflection (d$_{(004)}^{\textrm{ideal}}$ = 1.677 \angstrom; d$_{(004)}^{\textrm{turbo}}$ = 1.75 \angstrom ), which is 4.5\% larger than ideal. The increase in bond length is partially due to the presence of O trapped between sheets, whose concentration has been shown to correlate with the degree of stacking disorder \citep{joseph1983b,croat-MAPS2008}. Turbostratic graphite in LD Orgueil grains contains a higher concentration of O than their onion-like HD cousins, based upon EDXS measurements \citep{croat-MAPS2008}. Due to the difficulty of quantifying low-Z elements, however, EDXS analysis can only elucidate relative differences in O abundances, not absolute concentrations. Graphane, a hydrogenated form of graphene, is another possible cause of the larger interplanar spacings observed in G6 \citep{daulton2010}. In graphane, sheets of C atoms are buckled as H bonds to the sheet surfaces, which pulls C atoms out of the plane. However, graphane also exhibits a 5\% reduction in the hexagonal edge length relative to pure graphene because of sheet buckling. The other significant features in the SAED patterns are diffracted intensity peaks from the (100) and (110) reflections (d$_{(100)}^{\textrm{ideal}}$ = 2.131 \angstrom, d$_{(110)}^{\textrm{ideal}}$ = 1.231 \angstrom), which correspond to the hexagonal planar structure of graphene sheets. The measured d-spacings for these reflections are within 1\% of ideal graphene for both turbostratic and nanocrystalline regions. As we do not observe any edge-length contraction in the turbostratic or nanocrystalline planar structure, we therefore conclude that O, and not H, is most likely responsible for increased interplanar spacings.

\par The orientation of a crystal structure relative to the electron beam direction will determine whether the Bragg reflection conditions are satisfied, i.e. if the beam is parallel to the given crystal direction. When many crystals are randomly oriented about the beam axis, such as in powder diffraction, diffraction spots are distributed into rings. In microtomed sections of presolar graphite grains, graphite layers are oriented perpendicular to the section plane, so a SAED pattern will exhibit (002) rings with most of the diffracted intensity oriented in the direction where crystal layers are parallel to the beam (Figure \ref{fig:Ediff}).  Since the (002) crystal direction is orthogonal to both (100) and (110), the direction along which the maximum diffracted intensity lies from these reflections is perpendicular to the maxima in the (002) ring. There is a clear planar stacking direction in the turbostratic graphite, visible upon inspection and inferred from the diffraction patterns. In contrast, the diffracted intensity in the core SAED patterns is isotropic for all reflections (see Figures \ref{fig:Ediff}c and \ref{fig:DF}c,d), implying that the core consists of small randomly oriented graphene sheets. 

\par Unlike HD onion-type graphite grains with sharp boundaries between graphitic and nanocrystalline regimes, the transition in G6 occurs over a distance of 0.25 -- 0.5 \micron. SAED patterns obtained across this transition confirm a weakening (002) peak with decreasing radius. The (002) diffraction peak never disappears in the core, however, in contrast to what was observed in the cores of onion-type graphite grains \citep{bernatowicz1996,croat2005}, though it is less intense than the (100) and (110) reflections (Figure \ref{fig:Ediff}d). This implies that coherent planar stacking is present among some graphene sheets within the core. The measured (002) d-spacing is 4.9\% larger than ideal, similar to that of turbostratic graphite. All three microtome sections containing nanocrystalline material show the same (002) reflection feature, so it is unlikely to be due to a contribution from layers of turbostratic graphite included in the nanocrystalline region. Using the Scherrer formula D = 1.84$\lambda$/2$\Delta\Theta$, where D is the mass-weighted mean particle diameter, $\lambda$ is the wavelength of the electrons (0.0025 nm at 200 keV), and 2$\Delta\Theta$ is the angular FWHM of the diffraction peak, we estimate the lateral size of the nanocrystals to be 2--4 nm, similar to those obtained by \citep{bernatowicz1996,croat2005}, while the particle size predicted in the planar stacking direction (002) is $\sim$4.5 nm. Dark field (DF) images generated by using the diffracted intensity from the (002) reflection from the turbostratic graphite (Figures \ref{fig:DF}c-d) confirm that there are no large-scale graphitic features within the core which might have contributed to the presence of the small observed (002) reflection. There are, however, small intense spots in the DF images of the core, which may correspond to nanocrystals with planar stacking a few 10s of nm in size. These randomly-oriented ``larger'' nanocrystals may be responsible for the presence of the isotropic (002) reflection in SAED patterns of the core. 

\subsubsection{Subgrains}
\label{sec:subgrains}
\par In G6, numerous TiC subgrains (125--265 nm, 2550 ppm) were observed, as well as Fe and Fe-Ni subgrains (15--75 nm, 75 ppm), and one SiC subgrain (70 nm, 30 ppm). The TiC subgrains have V/Ti ratios between 0.098 and 0.104, in general agreement with previous observations of TiC subgrains in SN graphite grains \citep{croat2003}. TiC and VC are isostructural, allowing for V to exist in solid solution within the TiC structure. Si was the only other trace element observed in the TiC subgrains, with concentrations varying between 0.5 and 6.9 at.\%. The Fe-Ni subgrains have varying Ni/Fe ratios, between 0.023 and 0.253. The largest Fe-Ni grain (75 nm) is an intergrowth of two grains, the larger (65 nm) grain with a Ni/Fe ratio of 0.038, and the smaller (10 nm) grain with a Ni/Fe ratio of 0.253. The larger Fe-Ni grain also contains 1.3 $\pm$ 0.2 at.\% S, while the smaller grain does not contain any detectable S. The other Fe-Ni grains contain no detectable S, but do contain varying amounts of Si. One 28 nm grain with strong O and Si signals has a Ni/Fe ratio of 0.054 and a (Fe + Ni)/Si ratio of 1.9, resembling fayalitic olivine  containing Ni in solid solution with Fe, although we cannot determine the stoichiometry without a quantified O concentration. This grain was too small to index with electron diffraction. Notably, no subgrains were found within the nanocrystalline core; the TiC subgrain shown in Figure \ref{fig:Ediff}a lies at the core's edge.

\par
\begin{figure}[ht]
\figurenum{7}
\plotone{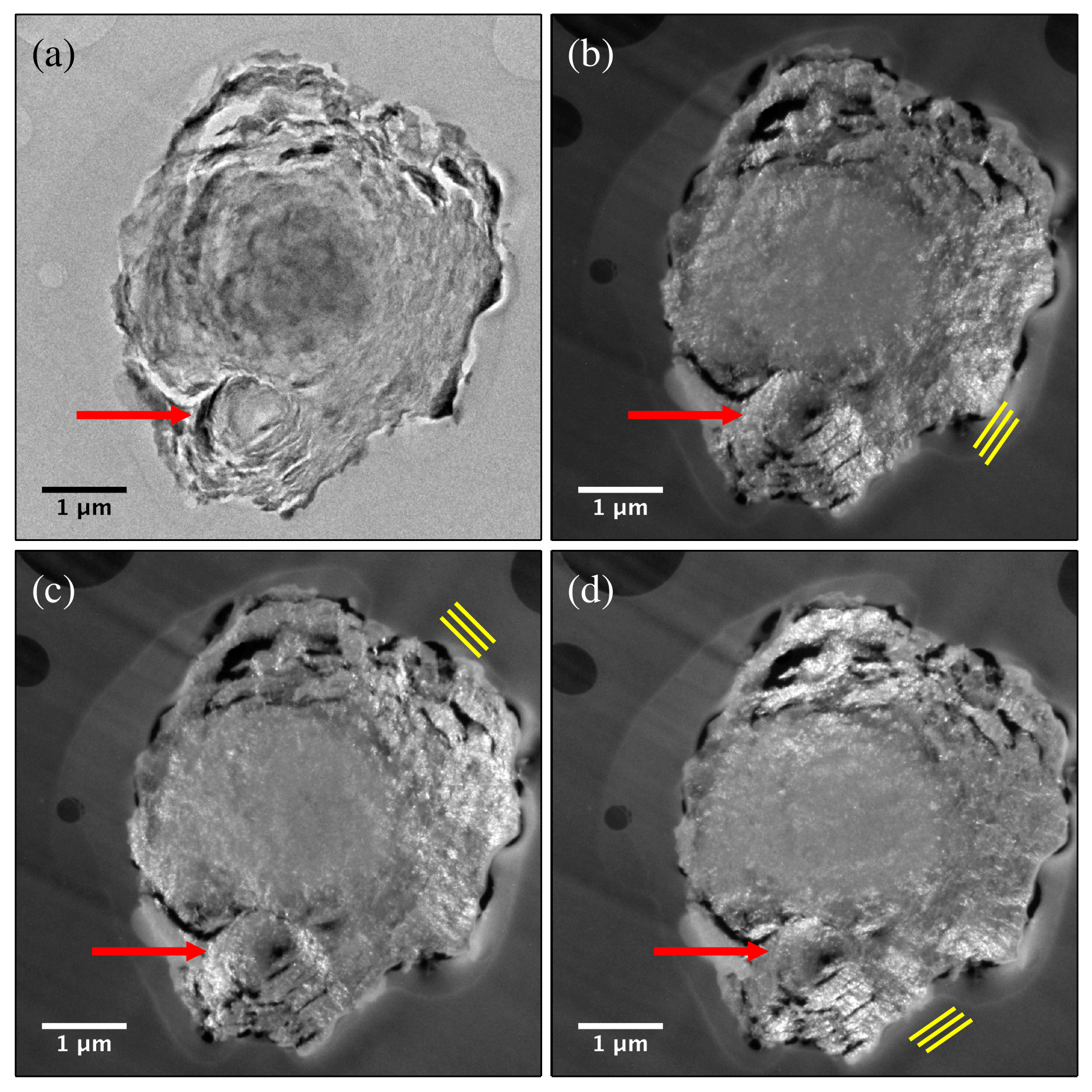}
\caption{\singlespace(a) TEM Bright field (BF) TEM micrograph of a section of presolar graphite grain G6. (b--d) Dark field (DF) images generated with diffracted electrons from the (002) reflection at three different angles (see Figure \ref{fig:DF}). G6's nanocrystalline core is visible in the center-left of the section. Concentric shells of turbostratic graphite, which make up the bulk of G6, clearly have two distinct centers: the largest shells are centered on the nanocrystalline core; directly below the core in the images is a second nucleation center around which smaller shells of turbostratic graphite formed (indicated by arrows). These likely existed as individual carbonaceous grains before they adhered to one another in the stellar ejecta. At the time of cohesion, G6 was predominantly nanocrystalline, while the smaller grain was turbostratic. Subsequently deposited turbostratic graphite grew around a joint center. Despite their coevolution, graphite subgrains within larger presolar graphite grains are rare \citep{croat-MAPS2008}.}
\label{fig:graphitesub}
\end{figure}

\begin{figure}[ht]
\figurenum{8}
\plotone{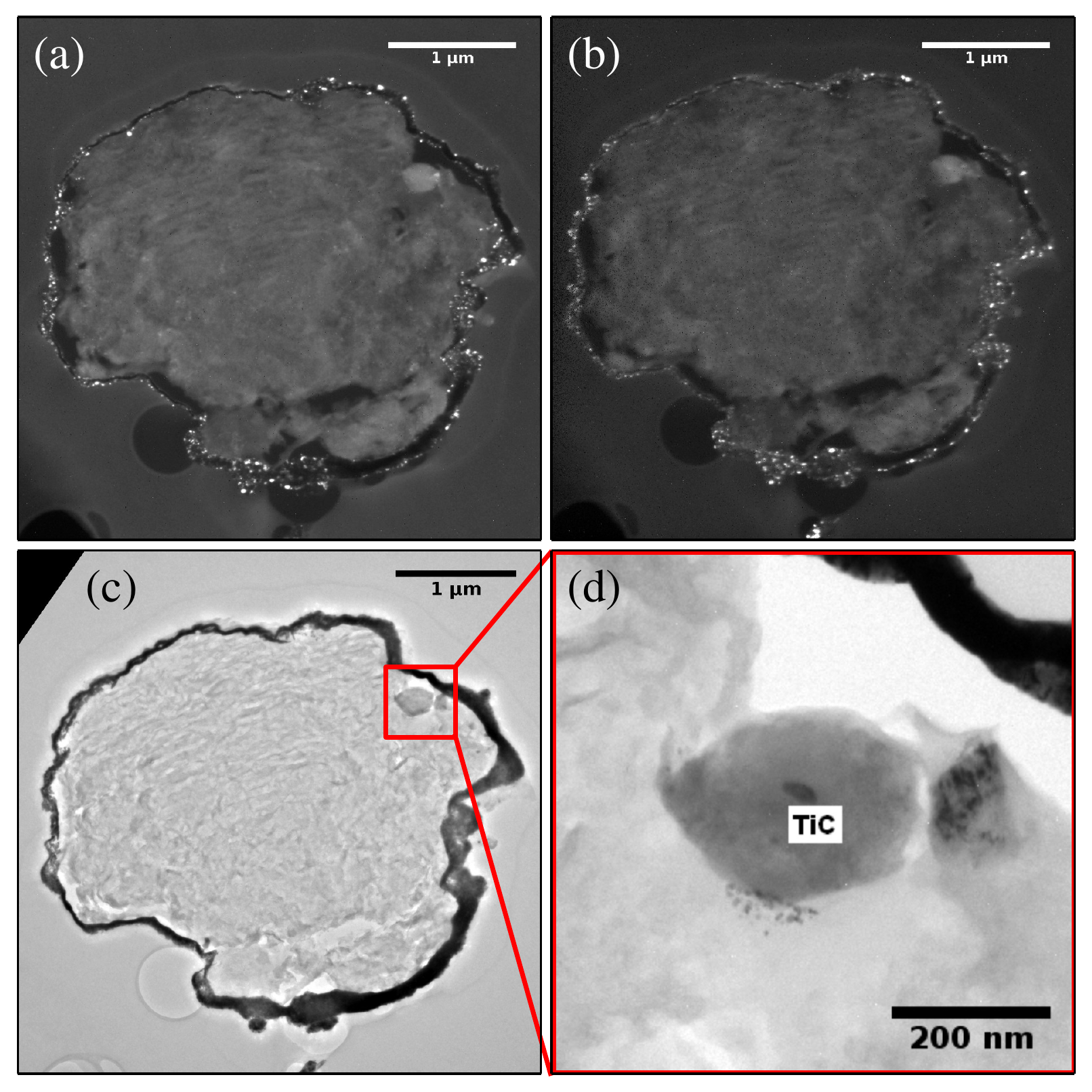}
\caption{\singlespace(a) and (b) Dark field (DF) images generated with diffracted electrons from the (110) and (100) reflections, respectively, of a section of G6, and (c) the corresponding bright field (BF) image. A small graphitic inclusion is clearly visible  in the DF images as its stacking direction is misaligned with the parent grain; it is also clearly visible due to diffraction contrast in BF imaging (boxed in c). Upon closer inspection, this graphite “subgrain” was found to contain a TiC subgrain near its center (d). The black rim around the section is redeposited Au from NanoSIMS analysis.}
\label{fig:inclusions}
\end{figure}

 We also observed two graphite spherule subgrains in G6. The larger one is nearly 1 \micron\ in diameter, with its own distinct nucleation center apparent via visual inspection and in DF images generated from (002) diffracted electron intensity (Figure \ref{fig:graphitesub}). This graphite subgrain became stuck to the larger grain when G6 was only two \micron\ in size and was mostly nanocrystalline. Turbostratic graphite layers continued to grow radially from the two different centers before the layers eventually unified around a common center. Despite the great abundance of metal and carbide subgrains within presolar graphite spherules, agglomerations of individual graphite grains remain quite rare, perhaps due to a low sticking probability between two comparably sized spherules. The second, smaller (300 nm) graphite subgrain is present at the surface of G6 and was clearly captured by the larger grain. This particular subgrain stands out in DF images formed with diffracted electron intensity from the (100) and (110) reflections (Figure \ref{fig:inclusions}). The graphite layers of this subgrain are clearly misaligned with the concentric shells of turbostratic graphite which comprise this section of G6. Additionally, this graphite subgrain contains its own 40 nm $\times$ 20 nm TiC subgrain at its center (Figure \ref{fig:inclusions}d).

\par G6 contains a unique microstructure relative to other presolar graphite grains. To date it remains the only LD SN graphite grain to contain a nanocrystalline core surrounded by a mantle of turbostratic graphite. G6's core is also unique vis-{\`a}-vis cores found in HD graphite grains, as it contains evidence for planar stacking of small, otherwise randomly-oriented 2-4 nm sheets of graphene. The microstructural heterogeneities within G6 provide further evidence for a complex and changing formation environment.

\subsection{Condensation}
\label{sec:condensation}
\par
\begin{figure}[ht]
\figurenum{9}
\plotone{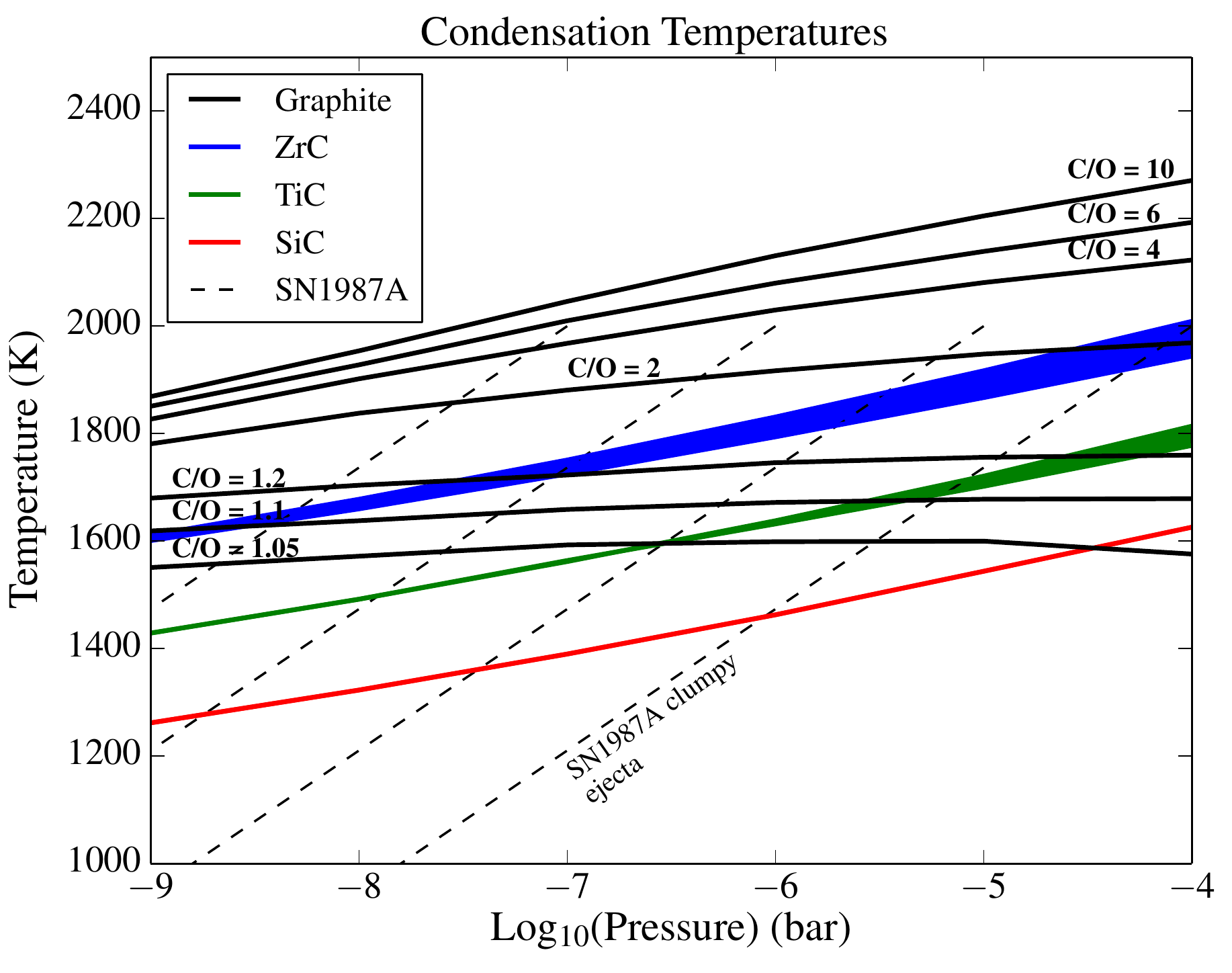}
\caption{\singlespace Condensation temperatures (T) for carbonaceous materials at different ambient pressures (P) and C/O ratios. The onset T for graphite condensation is highly dependent on the ambient C/O ratio, while the carbides are more sensitive to the ambient P. Envelopes for ZrC (blue in the online version) and TiC (green online) illustrate the range of T for a range of C/O ratios. Nanocrystalline conglomerates form when C is highly supersaturated, and potentially condense prior to any carbides. Carbides form next, which become trapped in turbostratic graphite condensing onto G6’s nanocrystalline core. SiC (red online) can only form prior to graphite at relatively high P. Dashed curves are P-T profiles that \citet{fedkin2010} derived (for mean atomic weight of 16) from \citet{wooden1993} observations of SN1987A's clumpy ejecta by reducing P by factors of 1, 10, 100, and 1000 (right to left). The P of SN1987A's clumpy ejecta dropped from $\sim$10$^{-4}$ bar on day 60 to $\sim$8$\times$10$^{-9}$ bar on day 777 (T = 500K, not shown). Adapted from \citep{bernatowicz1996} and \citep{fedkin2010}; condensation data from \citep{lodders1995}; SN1987A observations from \citep{wooden1993}. See the electronic edition for a color version of this figure.}
\label{fig:condensation}
\end{figure}

 The lack of subgrains within G6's nanocrystalline core is potentially significant for our understanding of the grain's formation sequence, though it is not statistically rigorous. Metal and carbide subgrains have been found within nanocrystalline cores of HD graphite grains (K. Croat, personal communication, December 2013). Numerical simulations \citep{cadwell1994} and laboratory observations \citep{bernatowicz1996,bernatowicz1997} have shown that refractory carbide grains, particularly TiC, can act as the nucleation point for graphite growth, which unambiguously imply that the TiC formed before graphite. The majority of TiC subgrains, however, have been found as inclusions, entrained as the graphite grain grew. The condensation temperatures, and therefore condensation sequence, for carbonaceous materials including graphite and metal carbides are highly sensitive to the ambient pressures and C/O ratios in stellar ejecta \citep{lodders1995,bernatowicz1996,fedkin2010} (Figure \ref{fig:condensation}). As the formation of a nanocrystalline conglomerate in lieu of the formation of graphite purportedly requires a high supersaturation of C \citep{bernatowicz1996}, the C partial pressure would also be high, yielding a large local C/O ratio. Since the system was out of equilibrium, this does not necessarily translate to a higher formation temperature. Instead, nanocrystalline conglomerates likely formed very quickly once nucleation points became available. As the saturation of C decreased, the system fell into thermodynamic equilibrium where the carbides were able to form, followed by the turbostratic graphite.
Recent laboratory investigations into the formation of carbon ``soot balls'' using PAH seeds \citep{contreras2013} may be applicable to the formation of nanocrystalline conglomerates. Using various carbonaceous seed molecules and extremely rapid cooling (temperature gradient of 3000K to 100K over a distance of 1.5--2.5 mm), \citet{contreras2013} ``freeze'' the plasma of seed molecules into conglomerates, whose end products can range up to microns in size. It should be noted, however, that this particular setup may not be ideal for the formation of carbonaceous particles with long-range crystalline order (100's of nm to 1 \micron), which under stellar conditions can take months \citep{hoppe2002} or years \citep{bernatowicz2005} to form.
\par Graphitization of carbonaceous materials has been found to decrease progressively with increasing oxidation \citep{joseph1983b}. Therefore, turbostratic graphite in LD grains likely required a higher relative O content in the parent gas than graphite found in HD grains, and so the C/O ratio ought to be closer to one in the SN ejecta where G6 formed compared to the AGB star ejecta that produced the HD grains. This is corroborated by recent studies of refractory metal nuggets (RMNs) found within HD graphite grains \citep{croat:MAPS2013}, which show that HD graphite grains form at higher temperatures than LD grains, which therefore requires a larger C/O ratio (see Figure \ref{fig:condensation}). This also implies a large temporal change in the ambient C/O ratio within the SN ejecta, from C supersaturation while the nanocrystalline core forms to C/O $\sim$ 1  when the turbostratic graphite mantle forms. HD grains, with near-ideal onion-like graphite condensing over their nanocrystalline cores, would require higher C/O ratios in order to avoid the crystallographic disruption of entrained O. Alternatively, CO dissociation via UV radiation \citep{clayton2013} might potentially increase the amount of free O available even at larger bulk C/O ratios, which might contribute to the formation of turbostratic graphite in SN ejecta. However, it seems unlikely that UV dissociation of CO molecules  would simultaneously not also disrupt C-C bonds. Additionally, the presence of a SiC subgrain near the surface of G6 implies that C/O $\sim$ 1 by the end of condensation.
\par Figure \ref{fig:condensation} also shows pressure and temperature (P-T) profiles \citep{fedkin2010} derived from observations of clumpy ejecta from SN1987A from 60 to 777 days post-explosion \citep{wooden1993}. The profiles were derived assuming equilibrium thermodynamics and an average atomic mass of 16 in the ejecta \citep{fedkin2010}. The ejecta's P-T profile evolves from P $\sim$ 10$^{-4}$ bar and T = 2000 K at day 60 to P = 8$\times$10$^{-9}$ bar and T = 500 K at day 777 (not shown in Figure \ref{fig:condensation}). Also plotted are P-T profiles for the ejecta with the pressure reduced by factors of 10, 100, and 1000. For clumpy SN ejecta similar to those of SN1987A, TiC would begin condensing at $\sim$1715 K and would only do so prior to graphite if the C/O ratio were less than $\sim$1.15. For lower-density ejecta, this formation sequence would be even more constrained. Based upon these observations and derived P-T profiles, it appears that presolar graphite grains from AGB stars and SNe may form under a similar set of physical  conditions. These profiles do not take into account the specific mixture of gas and its effect upon the condensation temperature.
\par Based upon equilibrium condensation and observations of SN 1987A, SN graphite grains are constrained to form in the first few months to one year post explosion. The interpretation of \iso{49}{Ti} excesses as originating from the decay of short-lived \iso{49}{V} ($t_{1/2}$ = 330 days) in some SiC X grains \cite{hoppe2002,lin2010} is in general agreement with this picture. Therefore the ejecta's chemical and isotopic changes as recorded in G6 must also have occurred on this timescale.

\subsection{XANES}
\label{sec:XANES}
\par XANES spectra from both the turbostratic graphite and nanocrystalline core regions of G6 exhibit graphite-like features (Figure \ref{fig:XANES}). The most dominant features include the \pistar\ absorption, located at 285.3 eV and 285.4 eV for the mantle and core, respectively; and the \sigstar\ absorption peaks located at 291.5 eV and 292.5 eV for both materials. The asymmetry parameters of the  \pistar\ peak shapes between the mantle and the core are the same. The core FWHM is 1.7 eV and the mantle FWHM is 1.9 eV, while the median energies of the peaks are 285.7 eV and 285.8 eV, respectively. It has been shown for polyaromatic hydrocarbons that as the domain size grows, i.e. as rings are added, discrete \pistar\ resonances around 285 eV begin to overlap and combine to form a single peak such as that observed in graphite \citep{brandes2008,cody1995,schwarz1987}. Additionally, decreasing domain size and amorphization of C smooth out the \sigstar\ peaks and reduce the intensity of the \pistar\ absorption \citep{gago2001}. The domain sizes of the core's nanocrystalline material, determined by electron diffraction, are large enough to retain strong \pistar\ and \sigstar\ absorbances. The primary difference between spectra of the mantle and core is that the mantle spectrum contains a carboxyl peak (288.6 eV) (Figure \ref{fig:XANES}a). This may be attributed to a higher concentration of O in the region where turbostratic graphite formed.  From this we can also infer that some portion of the O in the turbostratic region is strongly bonded to the disrupted graphite. From comparison of the carboxyl feature in the turbostratic graphite, which is apparent in all cross sections, to O isotope images detailed above, we can be confident that this peak is not due to a contaminant, as the turbostratic graphite contains the most O-anomalous regions within G6. The peaks at $\sim$288.6 eV from the deconvolved turbostratic and nanocrystalline spectra differ in magnitude by a factor of two, while the \pistar$_2$ peaks have the same height (Figures \ref{fig:XANES}b-c insets).

\section{Conclusions}
\label{sec:Conclusions}
\par Correlated NanoSIMS, TEM, and XANES studies of the Orgueil LD graphite grain G6 have revealed a wonderfully rich and unique SN graphite grain with a complex formation history. Isotopic and structural heterogeneities throughout the grain imply that G6's progenitor ejecta evolved chemically and isotopically over the timescale of months as the grain condensed. The magnitude and structure of isotopic and compositional heterogeneities leave little doubt that they are the product of primary formation and are not caused by secondary alteration or isotopic equilibration. To account for the observed O-anomalous shell, the O isotopic composition of G6's local environment would have had to change by a factor of 3--4 non-monotonically during its formation. These observations may provide evidence for fine-scale mixing in SN ejecta or a mechanism for grain transport through more coarsely mixed ejecta.

\acknowledgements
The authors would like to thank Jill Pasteris for providing a sample of Ceylon graphite, David Kilcoyne for assistance in operating the STXM at Beamline 5.3.2.2 at the Advanced Light Source, and Pat Gibbons and Kevin Croat for helpful discussions. This work was supported by NASA Earth and Space Sciences Fellowship (NESSF) NNX11AN60H, and NASA grants NNX10AI45G, NNX11AH14G, and NNX10AI63G. The Advanced Light Source is supported by the Director, Office of Science, Office of Basic Energy Sciences, of the U.S. Department of Energy under Contract No. DE-AC02-05CH11231. We would like to thank the reviewer for revisions and suggestions, which have improved this paper.

\pagebreak
\begin{deluxetable}{llllllrr}
\tabletypesize{\scriptsize}
\tablecaption{OR1d6m-6 isotopic composition. \label{table:G6}}
\tablewidth{0pt}
\tablehead{
\colhead{Ratio} &
\colhead{Bulk Grain\tablenotemark{\dag}} &
\colhead{Error ($1\sigma$)} &
\colhead{Average G6I} &
\colhead{Error ($1\sigma$)} &
\colhead{Terrestrial} &
\colhead{He/C Zone\tablenotemark{\dag\dag}} &
\colhead{He/N Zone\tablenotemark{\dag\dag}}
}
\startdata
\ratio{12}{13}{C} & \textbf{113} & 1     &\textbf{143}&1	& 89    & 35,000 & 4 \\
\delval{15}{14}{N} & \textbf{872} & 171     &{\bf38}	&18	& 0   & 9,100 & -990 \\
\delval{18}{16}{O} & \textbf{2057} & 74     &\textbf{1327}	&14	& 0   & 553,400 & -970 \\
\delval{29}{28}{Si} & \textbf{-60} & 16    &--&--& 0     & 307 & 1 \\
\delval{30}{28}{Si} & -13   & 13    &--&--& 0     & 342 & 0 \\
\delval{25}{24}{Mg} & 9 & 19 &--&--&0&143&-800\\
\delval{26}{24}{Mg} & \textbf{27,600} & 272 &--&--&0&893&462\\
\ratio{26}{27}{Al} & \textbf{4.46E-01} & 7.66E-03 &--&--& 0     & 0.02 & 0.2 \\
\delval{42}{40}{Ca} & \textbf{73} & 21    &--&--& 0     & 286 & -1 \\
\delval{43}{40}{Ca} & \textbf{108} & 46    &--&--& 0     & 264 & 1\ \\
\delval{44}{40}{Ca} & \textbf{36} & 14    &--&--& 0     & 19 & 3\ \\
\delval{46}{48}{Ti} & 39    & 26    &--&--& 0     & 44 & -1 \\
\delval{47}{48}{Ti} & 29    & 28    &--&--& 0     & -87 & -29 \\
\delval{49}{48}{Ti} & \textbf{467} & 40    &--&--& 0     & 1,900 & 7 \\
\delval{50}{48}{Ti}\tablenotemark{*} & \textbf{748} & 342   &--&--& 0     & 680 & 22 \\
\tableline
\enddata

\tablenotetext{\dag}{Bold values removed from normal by more than 2$\sigma$.}
\tablenotetext{\dag\dag}{Average zone composition from 15\Msun\ SN model \citep{rauscher2002}}
\tablenotetext{*}{60\% correction from \iso{50}{Cr} isobaric interference.}
\end{deluxetable}%
\pagebreak

\end{document}